\documentstyle[preprint,aps]{revtex}

\begin{document}
\preprint{ PUPT-1801 IASSNS-HEP-98-57 }
\title{Global Structure of Evaporating Black Holes}
\author{Maulik K. Parikh $^{a}$ \footnote{parikh@puhep1.princeton.edu}, 
Frank Wilczek $^{b}$ \footnote{wilczek@sns.ias.edu} }
\address{${}^{a \,}$ Joseph Henry Laboratories, 
Princeton University, Princeton, New Jersey 08544, USA}
\address{${}^{b \,}$ School of Natural Sciences, Institute for Advanced 
Study, Princeton, New Jersey 08540, USA}
\maketitle
\begin{abstract}
By extending the charged Vaidya metric to cover all of spacetime,
we obtain a Penrose diagram for the formation and evaporation of a charged
black hole. In this construction, the singularity is time-like. The
entire spacetime can be predicted from initial conditions if boundary
conditions at the singularity are known.

\end{abstract}

\def \pl {\partial}
\def \f {\frac}
\def \lf {\left (}
\def \rt {\right )}
\def \tu {\tilde u}
\def \del {\nabla}
\def \vaidya {\lf 1 - {2M(u) \over r} + {Q^2(u) \over r^2} \rt}

\section{Introduction}
It is challenging to envision a plausible global structure for a spacetime
containing a decaying black hole. If information is not lost in 
the process of black hole decay, then the final
state must be uniquely determined by the initial state, and vice versa.
Thus a post-evaporation space-like hypersurface must lie within the future
domain of dependence of a pre-evaporation Cauchy surface. One would like to
have models with this property that support approximate (apparent) horizons.

In addition, within the framework of general relativity,
one expects that singularities will form inside black holes \cite{singularity}.
If the singularities are time-like, one can
imagine that they will go over into the world-lines of additional degrees
of freedom occurring in a quantum theory of gravity.
Ignorance of the nature of these degrees of freedom is reflected in
the need to apply boundary conditions at such singularities.
(On the other hand, boundary conditions at future space-like singularities
represent constraints on the initial conditions; it is not obvious how a
more complete dynamical theory could replace them with something more
natural.)

In this paper,
we use the charged Vaidya metric to obtain a candidate macroscopic 
Penrose diagram for the
formation and subsequent evaporation of a charged black hole,
thereby illustrating how predictability might be retained. We do this
by first extending the charged Vaidya metric past its coordinate
singularities, and then joining together patches of spacetime that
describe different stages of the evolution.

\section{Extending the Charged Vaidya Metric}

The Vaidya metric \cite{vaidya} and its charged generalization
\cite{plebanskistachel,bonnorvaidya}
describe the spacetime geometry of unpolarized radiation, represented
by a null fluid, emerging from a spherically symmetric source.
In most applications, the physical relevance of the Vaidya 
metric is limited to the
spacetime outside a star, with a different metric describing the star's
internal structure.
But black hole radiance \cite{swh} suggests use
of the Vaidya metric to model back-reaction effects 
for evaporating black holes \cite{hiscockII,balbinot} 
all the way upto the singularity.

The line element of the charged Vaidya solution is
\begin{equation}
ds^2 = - \vaidya du^2 - 2 \, du \, dr 
+ r^2 \lf d \theta ^2 + \sin ^2 \theta d \phi ^2 \rt \; .	\label{bv}
\end{equation}
The mass function $M(u)$ is the mass measured at future null infinity 
(the Bondi mass) and is in general a decreasing function of the outgoing null
coordinate, $u$. Similarly, the function $Q(u)$ describes the charge, measured
again at future null infinity. When $M(u)$ and $Q(u)$ are constant, the
metric reduces to the stationary Reissner-Nordstr\"om metric.
The corresponding stress tensor describes a purely electric Coulomb field,
\begin{equation}
F_{ru} = + {Q(u) \over r^2} \; ,
\end{equation}
and a null fluid with current
\begin{equation}
k_a = k \del _a u \; , \; \; k^2 = +{1 \over 4 \pi r^2} 
{\pl \over \pl u} \lf - M + {Q^2 \over 2r} \rt \; .
\end{equation}
In particular,
\begin{equation}
T_{uu} = {1 \over 8 \pi r^2} \left [\vaidya {Q^2(u) \over r^2} + 
{1 \over r}{\pl Q^2 (u) \over \pl u} - 2 {\pl M (u) \over \pl u} \right ] \; . 
\label{vstress}
\end{equation}

Like the Reissner-Nordstr\"om metric, the charged Vaidya metric is beset
by coordinate singularities. It is not known how to remove these 
spurious singularities for arbitrary mass and charge functions 
(for example, see \cite{senovilla}).
We shall simply choose functions for which the relevant integrations
can be done and continuation past the spurious singularities can be carried
out, expecting that the qualitative structure we find is robust.

Specifically, we choose the mass to be a decreasing linear function
of $u$, and the charge to be proportional to the mass:
\begin{equation}
M(u) \equiv au + b \equiv \tu \; , \; \; Q(u) \equiv \eta \tu \; , 
\label{linear}
\end{equation}
where $a < 0$ and $|\eta| \leq 1$, with $|\eta| = 1$ at extremality.
We always have $\tu \geq 0$.
With these choices, we can find an ingoing (advanced time) 
null coordinate, $v$, with which the line
element can be written in a ``double-null'' form:
\begin{equation}
ds^2 = - {g(\tu,r) \over a} d \tu \, dv + r^2 \lf d \theta ^2 + 
\sin ^2 \theta d \phi ^2 \rt \; .
\end{equation}
Thus
\begin{equation}
dv = {1 \over g(\tu,r)} \left [ \lf 1 - {2 \tu \over r} + {\eta^2 \tu ^2 
\over r^2} \rt {d \tu \over a} + 2 \, dr \right ] \; .	\label{g}
\end{equation}
The term in brackets is of the form $X(\tu,r) \, d \tu + Y(\tu,r) \, dr$. 
Since $X(\tu,r)$ and $Y(\tu,r)$ are 
both homogeneous functions, Euler's relation provides the integrating factor:
$g(\tu,r) = X(\tu,r) \tu + Y(\tu,r) r$.
Hence
\begin{equation}
{\pl v \over \pl r} = \f{r^2}{r^3 +{\tu \over 2a}(r^2 -2 \tu r+ \eta^2 \tu ^2)}
\label{dvdr}
\end{equation}
\begin{equation}
{\pl v \over \pl \tu} = \f{ {1 \over 2a}(r^2 - 2 \tu r + \eta^2 \tu ^2) }
{ r^3 + {\tu \over 2a} (r^2 - 2 \tu r + \eta^2 \tu ^2 ) } \; .
\end{equation}
{}From the sign of the constant term of the cubic,
we know that there is at least one positive zero. 
Then, calling the largest positive zero $r'$, we may factorize the cubic as
$(r - r')(r^2 + \beta r + \gamma)$. Hence
\begin{equation}
\gamma = - {\eta^2 \tu ^3 \over 2 a r'} > 0 \; , \; \;
\gamma - \beta r' = - {\tu ^2 \over 2 a} > 0 \; , \; \;
\beta - r' = {\tu \over 2a} < 0 \; .
\end{equation}
Consequently, the cubic
can have either three positive roots, with possibly a double root but not
a triple root, or one positive and two complex (conjugate) roots. 
We consider these in turn.

{\it i) Three positive roots}

When there are three distinct positive roots, 
the solution to Eq. (\ref{dvdr}) is
\begin{equation}
v = A \ln (r - r') + B \ln (r- r_2) + C \ln (r - r_1) \; ,
\end{equation}
where $r' > r_2 > r_1 > 0$, and 
\begin{equation}
A = {+ {r'}^2 \over (r'- r_2)(r' - r_1)} > 0 \; , \; \;
B = {- {r_2}^2 \over (r' - r_2)(r_2 - r_1)} < 0 \; , \; \;
C = {+ {r_1}^2 \over (r' - r_1)(r_2 - r_1)} > 0 \; .
\end{equation}
We can push through the $r'$ singularity by defining a new coordinate,
\begin{equation}
V_2 (v) \equiv e ^{v / A} = (r-r') (r - r_2) ^{B / A} (r - r_1) ^{C / A} \; ,
\end{equation}
which is regular for $r > r_2$. 
To extend the coordinates beyond $r_2$ we define
\begin{equation}
V_1 (v) \equiv k_2 + (- V_2) ^ {A / B} 
= k_2 + (r' - r)^{A / B} (r_2 - r) (r - r_1)^{C / B} \; ,
\end{equation}
where $k_2$ is some constant chosen to match $V_2$ and $V_1$ at some 
$r' > r > r_2$. $V_1(r)$ is now regular for $r_2 > r > r_1$. 
Finally, we define yet another coordinate,
\begin{equation}
V(v) \equiv k_1 + (-(V_1 - k_2))^{B / C} 
= k_1 + (r' - r )^{A / C} (r_2 - r) ^{B / C} (r - r_1) \; ,
\end{equation}
which is now free of coordinate singularities for $r < r_2$.
A similar procedure can be applied if the cubic has a double root.

{\it ii) One positive root}

When there is only one positive root, $v$ is singular only at $r = r'$:
\begin{equation}
v = A \ln (r - r') + {1 \over 2} B \ln (r^2 + \beta r + \gamma )
+ \f { 2 C -  B \beta}{\sqrt{4 \gamma - \beta ^2}} 
\arctan \lf \f{2r + \beta}{\sqrt{4 \gamma - \beta ^2}} \rt \; .
\end{equation}
We can eliminate this coordinate singularity by introducing a new coordinate
\begin{equation}
V(v) \equiv e^{v/A} = (r - r') (r^2 + \beta r + \gamma)^{B / 2A}
\exp \left [+ {2 C - B \beta \over A \sqrt{4 \gamma - \beta ^2}} 
\arctan \lf \f{2r + \beta}{\sqrt{4 \gamma - \beta ^2}} \rt \right ]  \; ,
\label{V1zero}
\end{equation}
which is well-behaved everywhere. The metric now reads
\begin{equation}
ds^2 = -g(\tu, r){A \over V(\tu,r)} \, {d\tu \over a} \, dV+r^2 d \Omega^2 \; .
\end{equation}
In all cases, to determine the causal structure 
of the curvature singularity we express
$dV$ in terms of $du$ with $r$ held constant. Now we note that,
since $\tu$ is the only dimensionful parameter, all derived dimensionful
constants such as $r'$ must be proportional to powers of $\tu$. For example,
when there is only positive zero, Eq. (\ref{V1zero}) yields
\begin{equation}
\left . dV \right | _r = d \tu \, {V \over \tu} \left [ {-r' \over r - r'} 
+ {B \over 2 A} {\beta r + 2 \gamma \over r^2 + \beta r + \gamma} 
+ {2C - B \beta \over A \sqrt{4 \gamma - \beta^2}} 
{1 \over 1 + \lf \f{2r + \beta} {\sqrt{4 \gamma - \beta^2}} \rt ^2}
{-2r \over {\sqrt{4 \gamma - \beta ^2}}} \right ] \; .
\end{equation}
Thus, as $r \rightarrow 0$, and using the fact that $A + B = 1$, we have
\begin{equation}
ds^2 \rightarrow - {Q^2(u) \over r^2} \, du^2 \; ,
\end{equation}
so that the curvature singularity is time-like.

\section{Patches of Spacetime}

Our working hypothesis is that the Vaidya spacetime, since it incorporates
radiation from the shrinking black hole, offers a more realistic background
than the static Reissner spacetime, where all back-reaction is ignored.
In this spirit, we can model the black hole's evolution by joining
patches of the collapse and post-evaporation (Minkowski) phases onto
the Vaidya geometry.

To ensure that adjacent patches of spacetime match along 
their common boundaries, we can calculate the stress-tensor at their 
(light-like) junction. The absence of a stress-tensor intrinsic to 
the boundary indicates a smooth match when there is no explicit source there.
Surface stress tensors are ordinarily computed by applying junction 
conditions relating discontinuities in the extrinsic curvature;
the appropriate conditions for light-like shells were obtained 
in \cite{barisrael}. However, we 
can avoid computing most of the extrinsic curvature tensors by using the
Vaidya metric to describe the geometry on both sides of a given boundary,
because the Reissner-Nordstr\"om and Minkowski spacetimes are
both special cases of the Vaidya solution.

Initially then, we have a collapsing charged spherically symmetric 
light-like shell. Inside the shell, region I, the metric must be 
that of flat Minkowski space; outside, region II, it must be the 
Reissner-Nordstr\"om metric, at least initially.
In fact, we can describe both regions together by a 
time-reversed charged Vaidya metric,
\begin{equation}
ds^2 = - \lf 1 - {2M(v) \over r} + {Q^2(v) \over r^2} \rt dv^2 +2 \, dv \, dr +
r^2 d \Omega ^2 \; ,
\end{equation}
where the mass and charge functions are step functions of the ingoing null
coordinate:
\begin{equation}
M(v) = M_0 \Theta ( v - v_0 ) \; , \; \; 
Q(v) = \eta M(v) \; .
\end{equation}
The surface stress tensor, $t^s_{vv}$, follows from Eq. (\ref{vstress}). Thus
\begin{equation}
t^s_{vv} = {1 \over 4 \pi r^2} \lf M_0 - {Q^2_0 \over 2r} \rt \; .
\end{equation}
The shell, being light-like, is constrained to move at 45 degrees on
a conformal diagram until it has collapsed completely. 
Inside the shell, the spacetime is guaranteed by Birkhoff's theorem to
remain flat until the shell hits $r = 0$, at which point a singularity forms.

Meanwhile, outside the shell, we must have 
the Reissner-Nordstr\"om metric. This is appropriate for all $r > r_+$. 
Once the shell nears $r_+$, however, one expects that quantum effects 
start to play a role.
For non-extremal ($|\eta| < 1$) shells, the Killing vector changes
character -- time-like to space-like -- as the apparent horizon is 
traversed, outside the shell. This permits a virtual pair, created by a vacuum
fluctuation just outside
or just inside the apparent horizon, to materialize by having one member
of the pair tunnel across the apparent horizon. Thus, Hawking radiation
begins, and charge and energy will stream out from the black hole.

We shall model this patch of spacetime, region III, by the Vaidya metric. 
This must be attached to the Reissner metric, region II, 
infinitesimally outside $r = r_+$. A smooth match requires that there be 
no surface stress tensor intrinsic to the boundary of the two regions.
The Reissner metric can be smoothly matched to 
the radiating solution along the $u =0$ boundary if $b = M_0$ 
in Eq. (\ref{linear}).

Now, using Eqs. (\ref{g}) and (\ref{V1zero}), 
one can write the Vaidya metric as
\begin{equation}
ds ^2 = - {g^2(\tu,r) A \over \vaidya V^2 } \, dV^2 + 
2 {g(\tu,r) A \over \vaidya V} \, dV \, dr \; .
\end{equation}
We shall assume for convenience that $g(r)$ has only one positive 
real root, which
we call $r'$. Then, since $V$ and $g$ both contain a factor $(r - r')$, Eq.
(\ref{V1zero}), the
above line element and the coordinates are 
both well-defined for $r > r_+ (\tu)$. 
In particular, $r = \infty$ is part of the Vaidya spacetime patch.
Moreover, the only solution with $ds^2 = dr = 0$ also
has $dV = 0$, so that there are no light-like marginally trapped
surfaces analogous to the Reissner $r_{\pm}$. In other words,
the Vaidya metric extends to future null infinity, ${\cal I}^+$, and hence
there is neither an event horizon, nor a second time-like singularity on
the right of the conformal diagram.

The singularity on the left exists until the radiation
stops, at which point one has to join the Vaidya solution to 
Minkowski space. This 
is easy: both spacetimes are at once encompassed by a Vaidya solution with 
mass and charge functions
\begin{equation}
M(u) = (au + b) \Theta (u_0 - u) \; , \; \; Q(u) = \eta M(u) \; .
\end{equation}
As before, the stress tensor intrinsic to the boundary at $u_0$ can be read off
Eq. (\ref{vstress}):
\begin{equation}
t^s_{uu} = {1 \over 4 \pi r^2} \left [ (au+b)-{(au+b)^2 \over 2r} \right ] \; ,
\end{equation}
which is zero if $u_0 = - b/a$, i.e., if $\tu =0$. This says simply that
the black hole must have evaporated completely before one can return to
flat space.

Collecting all the constraints from the preceding paragraphs, we can
put together a possible conformal diagram, as in Fig. 1. 
(We say ``possible'' because a similar analysis for an uncharged hole 
leads to a space-like singularity; thus our analysis demonstrates 
the possibility, but not the inevitability,
of the behaviour displayed in Fig. 1.) 
Fig. 1 is a Penrose diagram showing the global structure of a spacetime
in which a charged imploding null shock wave collapses catastrophically 
to a point and subsequently evaporates completely. 
Here regions I and IV are flat Minkowski space,
region II is the stationary Reissner-Nordstr\"om spacetime, and region III
is our extended charged Vaidya solution. The zigzag line on the left 
represents the singularity, and the straight line separating region I
from regions II and III is the shell. The curve connecting the start of
the Hawking radiation to the end of the singularity is $r_+(\tu)$, which
can be thought of as a surface of pair creation. The part of region
III interior to this line might perhaps be better approximated by an
ingoing negative energy Vaidya metric.

{}From this cut-and-paste picture we see that, given some initial data set,
only regions I and II and part of region III can be
determined entirely; an outgoing ray starting at the bottom of the 
singularity marks the Cauchy horizon for these regions. 
Note also that there is no true horizon; the singularity is naked.
However, because the singularity is time-like,
Fig. 1 has the attractive feature that
predictability for the entire spacetime is restored 
if conditions at the singularity are known. It is tempting to
speculate that, with higher resolution, the time-like singularity 
might be resolvable into some dynamical Planck-scale object such as a D-brane.

{\bf Acknowledgement}

F.W. is supported in part by DOE grant DE-FG02-90ER-40542.

\begin{figure} [!hbt]
\input{epsf}
\epsfxsize = 5.0in \epsfbox{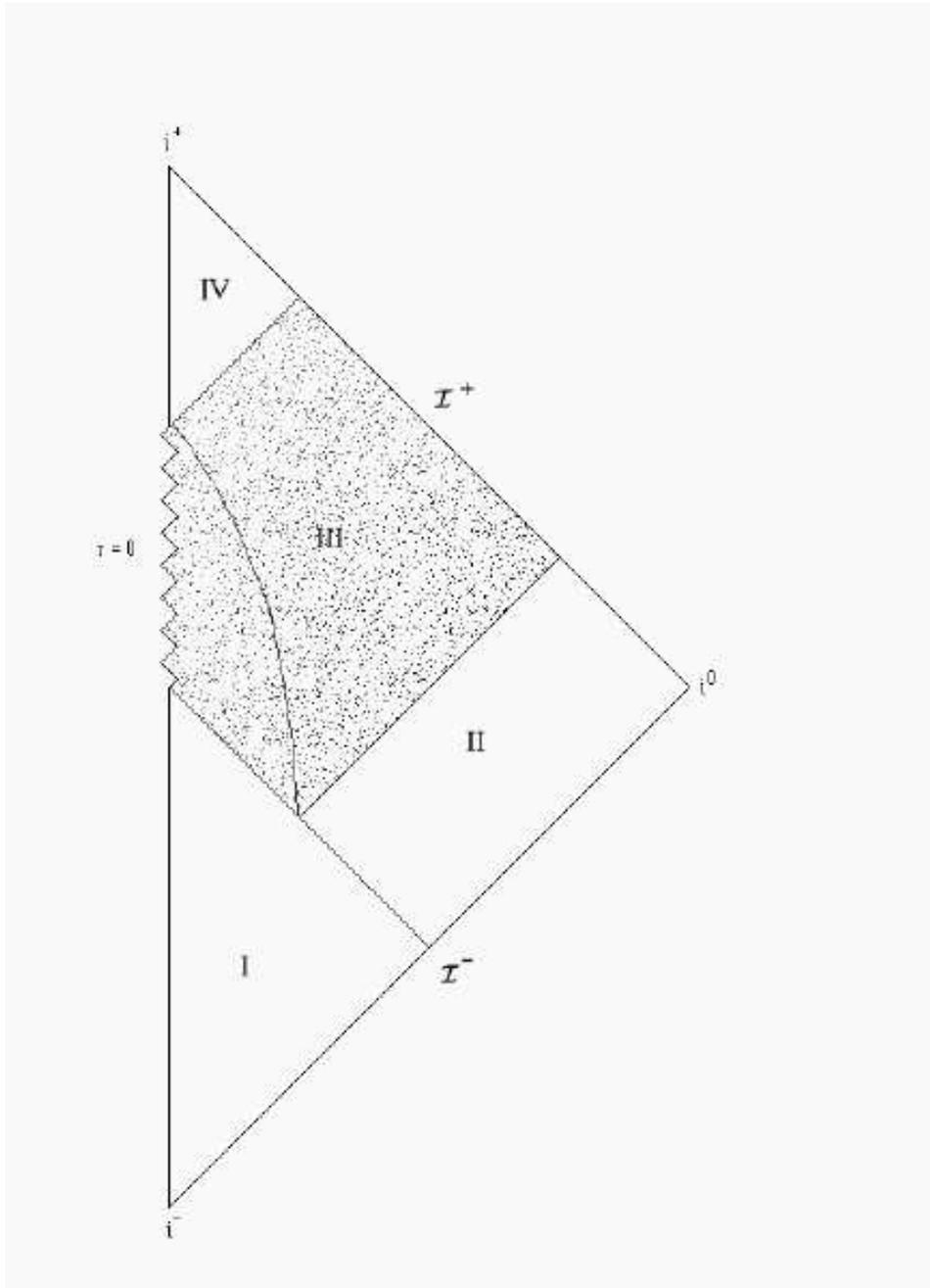}
\caption{Penrose diagram for the formation and evaporation of a charged black hole.}
\end{figure}

\end{document}